\documentclass[12pt,a4paper]{article}
\usepackage{amssymb}
\usepackage{amsmath}
\usepackage{epsfig}

\begin{document}
\title{Generalized Bloch Spheres for m-Qubit States.}
\author{Klaus Dietz\footnote{Permanent Address:Physics Department,University of
Bonn, 53115 Bonn, Germany, email: dietz@th.physik.uni-bonn.de}\\
Sektion Physik, LMU, Muenchen,\\ Theresienstrasse 37, 80333 Muenchen, and\\
\\Dipartimento di Scienze Fisiche ed Astronomiche \\dell'Universita di
Palermo,\\ via Archirafi 36, 90123 Palermo, Italy }
\maketitle
\begin{abstract}
m-Qubit states are embedded in $\mathfrak{Cl}_{2m}$
Clifford algebra. Their probability spectrum then depends on
$O\left(2m\right)$- or $O\left(2m+1\right)$-invariants respectively. Parameter
domains for $O\left(2m\left(+1\right)\right)$-vector and -tensor
configurations, generalizing the notion of a Bloch sphere, are derived. 
\end{abstract}
\section{Introduction}
For many purposes it is useful to consider m-qubit states as vectors in a 
$\mathbb{R}$-linear Hilbert space $\mathfrak{H}$ whose basis is a set
$\{B_i,\; i=1\ldots 2^{2m}\}$ of $2^m\times2^m$ orthonormal
\[
\text{trace}\left(B_i\cdot B_j\right)=\delta_{ij},
\]
hermitian matrices: 
\[
\mathfrak{H}=\{\sum_{k=1}^{2^{2m}}b_k\;B_k\;|\;b_k\;\;\text{real}\}.\qquad\left(H\right)
\]
A state is either represented by a hermitian, normalized matrix or an appropriate
coordinate vector $\left[b_1,b_2,\ldots,b_{2^{2m}}\right]$ (a formulation in an
appropriate \underline{projective} space would more adequate).In \cite{by}
\cite{ki} the generators of the quantum invariance group $SU\left(2^m\right)$
are proposed as such a basis, a possibility which we shall discuss in the Summary.\\
A straightforward solution for the
parametrisation of a state $\varrho$ (a density matrix) is to write the set of
all states as 
\begin{align*}
&\{\varrho\}=\bigcup_{\Lambda}\rho_{\Lambda}^{\mathfrak{U}}\\[1ex]
&\text{where}\nonumber\\
&\rho_{\Lambda}^{\mathfrak{U}}=\{U^{+}\varrho_{\Lambda} U\; |\;
U\in\mathfrak{U}\left(2^m\right)\}\nonumber\\[1ex]
&\text{and}\nonumber\\
&\text{$\varrho_{\Lambda}$ is the diagonal matrix}\nonumber\;\;
\varrho_{\Lambda}=\text{diag}\{\Lambda\},\nonumber\\[1ex]
&\Lambda=\{[\lambda_1,\lambda_2,\ldots,\lambda_{2^m}]\; |\;\lambda_i \text{
  real, }
\sum\lambda_i=1,\;\lambda_i\geq 0\};\nonumber
\end{align*}
$\mathfrak{U}\left(2^m\right)$ is the unitary group in $2^m$-dimensions and
$\Lambda$ is the probability spectrum generating the State
$\rho_{\Lambda}^{\mathfrak{U}}$. This construction warrants positivity and
normalisation. It is however not always (or,
better, almost never) convenient in the discussion of physical
situations.$\footnote{This is equally true for the parametrisation
  $\varrho=e^{A}/trace\left(e^{A}\right)$, $A$ hermitian, which is rather
  clumsy e.g. when it comes to the discussion of separability conditions.}$ On the
other hand writing $\varrho$ as a vector in $\mathfrak{H}$ 
confronts us with the problem of deriving conditions for the expansion
coefficients $\footnote{These parameters are linearly related, a matrix
  representation of the basis in $\mathfrak{H}$ $\quad B_i, i=1\ldots 2^{2m}$
  given, to the matrix elements of the density matrix.}$ that guarantee the
expansion to yield a state. Formulated in this 
general way the problem has no obvious solution: positivity and normalisation
conditions can derived by expressing the eigenvalues in terms of the expansion
coefficients, i.e. finding the zeroes of the characteristic polynomial as
functions of these parameters. As we know from Abel and Galois a solution by
rational operations and radicals does not exist for quintic or higher degrees,
i.e. for general 3-qubit and a fortiori for higher systems. For the 2-qubit
explicit expressions are given by the Ferrari-Cardano formulae.\\ In this
paper I explicitly construct classes of states for all $m$ whose 
spectra are determined by charactistic polynomials factorizing into
polynomials of a given degree. The novel point in our considerations is the use of
 {\sl hermitian matrix representations of a Clifford algebra} to construct bases in 
$\mathfrak{H}$. This particular choice of basis allows us to arrange the
$2^{2m}-1$ real coordinates of a m-Qubit state in multidimensional arrays
which are shown to 'transform' as $O\left(2m\right)$ tensors. This fact
implies that the probability spectrum of a m-Qubit state depends only on
$O\left(2m\right)$-invariants, a considerable simplification of the parameter
dependencies of these eigenvalues, indeed. This simplification leads to a
complete characterisation of complete$\footnote{Complete in the sense that 
all states factorizing in a specific way are contained in this set.}$ sets of
states which allow for an explicit construction of a parameter domain. In this
way I find the set of all states (vector-states) whose parameter domain is the Bloch
$2m$-sphere. Furthermore a set of (bivector)-states is proposed whose
novel parameter domain generalizes the notion of a Bloch sphere. Beyond
these two domains the Descartes rule for the positivity of polynomial roots
can be used to derive admissible parameter domains.
      
\section{m-Qubit states imbedded in Clifford Algebras.}

An m-qubit system is controlled by $m$ spin-degrees of freedom and hence by
$2^{2m}-1$ parameters (see footnote 2 on page 2). The determining anticommutation relation for
Clifford numbers \cite{cliff} ($\mathbb{I}$ is the unity)
\begin{eqnarray}
&&\Gamma_{i}\cdot\Gamma_{j}\;+\;\Gamma_{j}\cdot\Gamma_{i}=2\delta_{ij}\;\mathbb{I}\\[2ex]
&&\text{with}\nonumber\\[1ex]
&&i,j=1 \ldots 2m\nonumber
\end{eqnarray}
has $2^m$-dimensional, hermitian, traceless matrix representations
$\Gamma^{\{m\}}_{j}$.\\[1ex] From the anticommutation relations we see
immediately that the products
\begin{align}
&\Gamma_{j_1,j_2,\ldots,j_k}:=i^{k-1}\;\Gamma_{j_1}\cdot\Gamma_{j_2}\cdot\;\ldots\;\cdot\Gamma_{j_k}\\[2ex]
&k=2\ldots m
\end{align}
are totally anti-symmetric in the indices $\left[j_1\ldots j_k\right]$. 
The only symmetric object constructed from Clifford numbers is the unity
\[
 \mathbb{I}=\Gamma_i^2
\]
as we see from the anticommutation relations. A product consists of at most
$2m$ factors. Hence  we have 
\[
\sum_{k=0}^{2m}\binom{2m}{k}=2^{2m}
\]
independent products. Furthermore because of the commutation relations we have
\begin{align*}
&\text{trace}\left(\left(\Gamma^{\{m\}}_{i_1}\cdot\Gamma^{\{m\}}_{j_1}\cdot\;\ldots\;\cdot\Gamma^{\{m\}}_{k_1}\right)^{+}
\Gamma^{\{m\}}_{i_2}\cdot\Gamma^{\{m\}}_{j_2}\cdot\;\ldots\;\cdot\Gamma^{\{m\}}_{k_2}\right)\sim\\[1ex]
&\sum\left(\delta_{\bar{i}_1\bar{i}_2}
\delta_{\bar{j}_1\bar{j}_2}\ldots\delta_{\bar{k}_1\bar{k}_2}\right),\\[1ex]
&\text{where the $\delta$-function expresses pairwise equality of the
  $\cdot_1$- and $\cdot_2$-indices}.
\end{align*}
A hermitian $2^m\times 2^m$-matrix requires $2^{2m}$ real numbers for a
complete para-metrisation. Thus m-qubit states can be expanded in terms of
$\mathbb{I}$ and the products introduced: Clifford numbers are the starting
point for the  construction of a basis in the $\mathbb{R}$-linear space of
hermitian matrices:\\ this basis is construed as a Clifford algebra
$\mathfrak{Cl}_{2m}$ ($2^{2m}$-dimensional as we have seen). The important
advantage to gain from this choice of basis is that now domains for parameters
are determined by $O\left(2m\right)$-invariants. The number of parameters
necessary for the specification of these domains is thus considerably
reduced. For the domains found in this
paper this means one invariant for the vector-state configuration ($2m$
parameters) and two invariants for the bivector states ($m\left(2m-1\right)$ parameters)
to be constructed below for all $m$. \\  I should remark that many beautiful geometric
reverberations of Clifford algebras will play no r\^ole here, only very
elementary properties of Clifford algebras will be sketched, emphasizing
practical aspects. It is in this sense that the following, hopefully
selfcontained, outline of the method should be understood.\\ To construct a
basis and its matrix representation $\mathfrak{G}^{\{m\}}$ in $\mathfrak{H}$
\\ I proceed as follows:      
\vskip .3cm
\begin{itemize}
\item The product
\begin{equation}
\Gamma^{\{m\}}_{2m+1}:=\left(-i\right)^{m}\;\;\Gamma^{\{m\}}_{1}.
\Gamma^{\{m\}}_{2}.\ldots\Gamma^{\{m\}}_{2m}
\end{equation}
obviously anti-commutes with all the $\Gamma^{\{m\}}_{i}\;\,i=1\ldots 2m$.
\item The explicitly anti-symmetric products ($\varepsilon$ is the totally
  anti-symmetric symbol in $2m$-dimensions)
\begin{align}
&\hat{\Gamma}^{\{m,k\}}_{i_1\ldots i_k}=F_{Norm}^{\{m,k\}}\;\left(\varepsilon_{i_1\ldots
  i_{2m}}\;\Gamma^{\{m\}}_{i_{k+1}}.\ldots\Gamma^{\{m\}}_{2m}\right)\Gamma^{\{m\}}_{2m+1}\\[1ex]
&F_{Norm}^{\{m,k\}}=\frac{\left(-i\right)^{m+s}}{\left(2m\right)!}\binom{2m}{k}\nonumber\\[1ex]
&s=\begin{cases}
0\;\,&\text{when }\;x=0,1\\
1\;\,&\text{when }\;x=2,3
\end{cases}\quad\text{where }\;\,x=k\,mod\left(4\right)
\end{align}
The limiting cases $k=1$ and $k=2m$ are immediately seen to be
\begin{align*}
&\hat{\Gamma}^{\{m,1\}}_{i_1}\;=\;\left(-1\right)^{m+1}\Gamma^{\{m\}}_{i_1}\\[2ex]
&\hat{\Gamma}^{\{m,2m\}}_{i_1\ldots i_{2m}}=\frac{f_m}{2m!}\;\varepsilon_{i_1\ldots
  i_{2m}}\;\Gamma^{\{m\}}_{2m+1}\\[2ex]
&\text{with}\\[1ex]
&f_m =\begin{cases}
\left(-1\right)^\frac{m}{2}&\quad\text{for m even}\\
\left(-1\right)^{\frac{m+1}{2}}&\quad\text{for m odd}
\end{cases}
\end{align*}
\item Because of the anti-commutation relations the only symmetric tensor is
  the scalar, i.e. the unit matrix
\begin{align}
&\hat{\Gamma}^{\{m,0\}}\;=\;\mathbb{I}
\end{align}
\item The set of matrices
  ${\mathfrak{G}}^{\{m\}}\;=\;\{\hat{\Gamma}^{\{m,0\}},\hat{\Gamma}^{\{m,1\}}\;\;\ldots\;\;\hat{\Gamma}^{\{m,2m\}}\}$
  is orthonormal in the sense of (1).
\item Formally speaking this gives an identification of the linear spaces\\
$\mathfrak{g}^{\{m,k\}}=\text{span}\left(\hat{\Gamma}^{\{m,k\}},\mathbb{R}\right)$ and the
    tensor algebra $\overset{k}{\bigwedge}\;\mathbb{R}^{2m}$ of $\mathbb{R}^{2m}$. In
    detail we write$\footnote{We use the slightly old
  fashioned notation: vector,tensor,...k-tensor instead of vector, bivector,...k-vector}$ 
\vspace{2mm}
\begin{align*}
\text{Isomorphic vector spaces:}\\[2ex]
\text{scalar},\;\,\mathbb{R}\qquad&|\qquad\mathfrak{g}^{\{m,0\}}=\mathbb{R}\cdot
\mathbb{I}\\[2ex]
\text{vector},\;\,\overset{1}{\bigwedge}\;\mathbb{R}^{2m}=\mathbb{R}^{2m}\qquad&|\qquad\hat{\mathfrak{g}}^{\{m,1\}}\\[2ex]
\text{(2-)tensor
  (bivector)},\;\,\overset{2}{\bigwedge}\;\mathbb{R}^{2m}\qquad&|\qquad\hat{\mathfrak{g}}^{\{m,2\}}\\[2ex]
\ldots&\ldots\\[2ex]
\text{volume element},\;\,\overset{2m}{\bigwedge}\;\mathbb{R}^{2m}\qquad&|\qquad\hat{\mathfrak{g}}^{\{m,2m\}}\\
\end{align*}
\item Following these observations we organize the state-parameters in terms
  of a scalar $G_{0}^{\{m,0\}}$ and the totally anti-symmetric real arrays 
  \begin{align*}
&G_{i_1}^{\{m,1\}},\;G_{i_1,i_2}^{\{m,2\}}\;\ldots\;G_{i_1,i_2,\ldots
  ,i_{2m}}^{\{m,2m\}}\\[1ex]
&\text{(i.e. totally antisymmetric arrays of real numbers)} 
\end{align*}
and thus account for
\begin{equation}
\sum_{k=0}^{2m}\binom{2m}{k}\;=\;2^{2m}
\end{equation}
coefficients.
\item We write the expansion of a m-qubit state as 
\begin{equation}
\varrho^{\{m\}}\;=\;\sum_{k=0}^{2m}G^{\{m,k\}}\circ\hat{\Gamma}^{\{m,k\}}
\end{equation}
where $\circ$ indicates the contraction
$A\circ B=\sum_{i_1,\ldots,i_k}A_{i_1,\ldots,i_k}B_{i_1,\ldots,i_k}$.
\item An explicit construction of the representation
  $\Gamma^{\{m\}}_1,\ldots,\Gamma^{\{m\}}_{2m}$ traditionally proceeds e.g. as
  follows:\\ Starting with the Pauli matrices
\[
\sigma_1=\begin{pmatrix}0&1\\1&0 \end{pmatrix},\quad 
\sigma_2=\begin{pmatrix}0&-I\\I&0\end{pmatrix},\quad
\sigma_3=\begin{pmatrix}1&0\\0&-1\end{pmatrix},\quad
\sigma_4=\begin{pmatrix}1&0\\0&1\end{pmatrix},\quad
\]
we have the iteration scheme
\begin{align}
&{\mathfrak{G}}^{\{m+1\}}\;=\nonumber\\[1ex]
&\{\Gamma^{\{m,1\}}\times\sigma_1,\ldots,\Gamma^{\{m,2m\}}\times\sigma_1,\Gamma^{\{m,0\}}\times
\sigma_2,\;\,\Gamma^{\{m,0\}}\times\sigma_3\}\\\nonumber
\end{align}
\item $O\left(2m\right)$-symmetry:\\[1ex] To begin with it might be useful to
  remind the reader the machinery of rotations in classical systems. Consider
  a canonical, classical system with $2m$ degrees of freedom, i.e. with a
  $2m$-dimensional configuration space. Infinitesimal $2m$-dimensional
  rotations and translations generated by generators 
\[
J_{i,j}\,,\;P_i\quad\text{ respectively}
\]
($\{A\;,B\}$ denote Poisson brackets
  for functions defined on the phase space of the system) are defined as 
\begin{align*}
&\text{Infinitesimally:}\\[1ex]
&F\;\longrightarrow\; F+\epsilon_1\;\alpha_{i,j}\{J_{i,j}\;,F\}+\epsilon_2\;\beta_{i}\{P_i\;,F\}\\[1ex]
&\text{(repeated indices are summed over)}\\[2ex]
&\text{where}\\[1ex]
&\epsilon \text{ is infinitesimal and } \alpha_{i,j}\;\; i,j=1..2m\\
& \text{is an antisymmetric array of parameters}\\
&\text{the }\beta_i\text{ parametrize translations}.
\end{align*}
The Lie algebra of the Euclidean Poincar\'e group
\begin{align*} 
&\{J_{i,j}\;,J_{k,l}\}=\delta_{i,l}J_{j,k}+\delta_{j,k}J_{i,l}-\delta_{i,k}J_{j,l}-\delta_{j,l}J_{i,k}\\[1ex]
&\{J_{i,j}\;,P_k\}=P_i\delta_{j,k}-P_j\delta_{i,k}.
\end{align*}
The anticommutation relations (1) defining the Clifford algebra
$\mathfrak{Cl}_{2m}$ spanned by the set of totally antisymmetric products and
the unity $\mathfrak{G}=\{\mathbb{I},\Gamma_i,i\Gamma_i \Gamma_j,\ldots\}$
considered above lead to an analogous algebraic structure. A straightforward
calculation shows 
($\Gamma_{i,j}:=i\Gamma_i\cdot\Gamma_j$) 
\begin{align}
&\frac{i}{2}\left[\Gamma_{i,j}\;,\Gamma_{k,l}\right]=\delta_{i,l}\Gamma_{j,k}+\delta_{k,j}\Gamma_{i,l}-\delta_{l,j}\Gamma_{i,k}
-\delta_{i,k}\Gamma_{j,l}\\[1ex]
&\frac{i}{2}\left[\Gamma_{i,j}\;,\Gamma_k\right]=\delta_{k,i}\Gamma_j
-\delta_{j,k}\Gamma_i .
\end{align}
These relations constitute a quantum analogue of the classical
representation of the 
$O\left(2m\right)$ Lie algebra$\footnote{Precisions concerning a more precise
  discussion of the universal covering group are of no avail here and will not
be touched.}$: the $\Gamma_{i,j}$ generate rotations, the
$\Gamma_i$ translations in the Clifford algebra $\mathfrak{Cl}_{2m}$, the array
$\{\Gamma_1,\Gamma_2,\dots,\Gamma_{2m}\}$ 'transforms as a vector'. The basis elements
of the dual Grassmann algebra $\bigwedge\mathbb{R}^{2^m}$ can be identified with (see
above) \\ $\mathbb{G}=\{G^{\{m,0\}},G^{\{m,1\}},\ldots,G^{\{m,2m\}}\}$ and
'transform as tensors'. More precisely we have
\begin{align}
&L\in O\left(2m\right)\longmapsto
U\left(L\right)=e^{-\frac{i}{4}\alpha_{i,j}\Gamma_{i,j}}\\[2ex]
&O\left(2m\right)\text{-transformations}\nonumber\\
&G^{\{m,1\}}_i\longmapsto L_{i,k}G^{\{m,1\}}_k\\
&G^{\{m,2\}}_{i,j}\longmapsto L_{i,i_1}L_{j,j_1}G^{\{m,2\}}_{i_1,j_1}\\
&etc\nonumber\\
&\text{induce transformations}\nonumber\\[2ex]
&\Gamma_i\longmapsto U\left(L\right)\Gamma_i U\left(L\right)^{-1}=\left(L^{-1}\right)_{i,k}\Gamma_k\\
&\Gamma_i\Gamma_j\longmapsto U\left(L\right)\Gamma_i\Gamma_j
U\left(L\right)^{-1}=
\left(L^{-1}\right)_{i,i_1}\left(L^{-1}\right)_{k,k_1}\Gamma_{i_1}\Gamma_{k_1}\\
&etc.
\end{align}
Configurations parametrized by one of the tensors $G^{\{m,k\}}$ have 
some comfortable (and profitable) properties. For instance the coefficients of
the characteristic polynomials are expected to depend on
$O\left(2m\right)$-invariants built from these tensors. Furthermore the
probability spectra will exhibit degeneracy patterns corresponding to the rank
of the tensors $G^{\{m,k\}}$, parameter ranges corresponding to physical
states will be determined by universal polynomials in terms of these
invariants.  
\end{itemize}
The following sections are devoted to detailed dicussions of these
observations for the cases of m=2,3-qubits. General results for m-qubits will
be derived.

\section{$O\left(2^m\right)$-Tensor
  Configurations}
 
In this chapter I introduce some nomenclature which derives from similar
objects ocurring in the Dirac theory of relativistic Fermions.\\ 
The iteration scheme (10) provides us with explicit bases for Clifford
algebras $\mathfrak{Cl}_{2m}$.\\ The coordinates representing a m-Qubit
introduced in equation (H) of the Introduction are organized in 
\begin{itemize}
\item scalar $G^{\{m,0\}}$, $G^{\{m,0\}}=1$ because of state normalisation 
\item vector $G^{\{m,1\}}$,
\item 2,3-tensor $G^{\{m,2,3\}}$, and
\item pseudoscalar $G^{\{m,2m\}}$,
\item pseudovector $G^{\{m,2m-1\}}$,
\item pseudotensor $G^{\{m,2m-\left(2,3\right)\}}$
\end{itemize}
 components. $\footnote{Here we follow the nomenclature of Dirac
  theory (generalized for $m\neq 2$) for relativistic fermions choosing a
  euclidean Majorana representation for $\hat{\Gamma}^{\{m,k\}}$ generated by the
  iteration scheme (10).}$
\vskip.3cm
\begin{itemize}
\item ${\bf m=1}$\\[1ex] The 2-Clifford algebra is spanned by $\footnote{We could have chosen
\begin{eqnarray}
&&\hat{\Gamma}^{\{1,0\}}=\sigma_4\nonumber\\[1ex]
&&\hat{\Gamma}^{\{1,1\}}=\{\sigma_2,\,\sigma_3\},\; \text{or } \{\sigma_3,\sigma_1\}\\[1ex]
&&\hat{\Gamma}^{\{1,2\}}=\sigma_1,\;\text{or }\sigma_2\nonumber
\end{eqnarray} 
as well . Both basis are
connected by an $O\left(2\right)$ rotation by $\pi/4$.}$
\begin{eqnarray}
&&\hat{\Gamma}^{\{1,0\}}=\sigma_4\qquad\quad\quad\;\;\text{scalar}\nonumber\\[1ex]
&&\hat{\Gamma}^{\{1,1\}}=\{\sigma_1,\,\sigma_2\}\qquad\text{vector}\\[1ex]
&&\hat{\Gamma}^{\{1,2\}}=\sigma_3\qquad\qquad\;\;\text{pseudoscalar}\nonumber
\end{eqnarray}
A qubit state is then written as ($G^{\{m,o\}}=\frac{1}{2^m}$ because of
normalisation) 
\begin{equation}
\varrho=\frac{1}{2}\left(G^{\{1,0\}}\hat{\Gamma}^{\{1,0\}}\,+G^{\{1,1\}}\circ\hat{\Gamma}^{\{1,1\}}\,+
\,G^{\{1,2\}}\;\hat{\Gamma}^{\{1,2\}}\right)
\end{equation}
\vskip.3cm
\item ${\bf m=2}$\\[1ex] The Clifford algebra is now spanned by  
\[
\hat{\Gamma}^{\{2,1\}}_1=\begin{pmatrix}0&0&0&1\\0&0&1&0\\0&1&0&0\\1&0&0&0\end{pmatrix}\quad
\hat{\Gamma}^{\{2,1\}}_2=\begin{pmatrix}0&0&0&-i\\0&0&i&0\\0&-i&0&0\\i&0&0&0\end{pmatrix}\quad
\]
\begin{equation}
\end{equation}
\[
\hat{\Gamma}^{\{2,1\}}_3=\begin{pmatrix}0&0&i&0\\0&0&0&i\\-i&0&0&0\\0&-i&0&0\end{pmatrix}\quad
\hat{\Gamma}^{\{2,1\}}_4=\begin{pmatrix}1&0&0&0\\0&1&0&0\\0&0&-1&0\\0&0&0&-1\end{pmatrix}.
\]
We write
\begin{align}
&\varrho=\\[1ex]
&\frac{1}{4}\left(G^{\{2,0\}}\hat{\Gamma}^{\{2,0\}}\,+\,G^{\{2,1\}}\circ\hat{\Gamma}^{\{2,1\}}\,+\,
  G^{\{2,2\}}\circ\hat{\Gamma}^{\{2,2\}}\,+\,
  G^{\{2,3\}}\circ\hat{\Gamma}^{\{2,3\}}\right.\nonumber\\[1ex]
&+\left.G^{\{2,4\}}\;\hat{\Gamma}^{\{2,4\}}\right)\nonumber
\end{align}
\end{itemize}
The iteration algorithm (10) straightforwardly provides analogous representations for $m>2$.

\section{Probability Spectra for Tensor Configurations and Their
  Degeneracies.}

In this section we explicitly determine the $m=1,2,3$ probability spectra of
the vector and tensor configurations by calculating the roots of the
characteristic polynomial
\[
P^{\{m\}}:=\;\text{Determinant}\left(\varrho^{\{m\}}-\lambda\mathbb{I}\right);
\]
parameter domains generalizing Bloch
spheres are obtained by requiring that the spectrum obtained be a probability
distribution. Degeneracies of m-qubit tensor spectra are shown to 
follow simple patterns. Because of the normalisation condition a 'tensor
configuration' always reads as
\begin{align}
&\varrho_{k_{tensor}}=\frac{1}{2^m}\left(\mathbb{I}_m+G^{\{m,k_{tensor}\}}\circ\hat{\Gamma}^{\{m,k_{tensor}\}}\right)
\end{align}
We find 
\begin{itemize}
\item\underline{Vector configurations}: $k_{tensor}=1$\\[1ex]
The probability spectra are $2^{m-1}$-fold degenerate, i.e. built up by one
doublet repeated $2^{m-1}$-times. The doublet is found to be
\begin{align}
\lambda =\frac{1}{2^m}\left(1\pm \|G^{\{m,1\}}\|\right)
\end{align}
where the absolute value of $G^{\{m,1\}}$ is simply the vector norm
\begin{align}
\|G^{\{m,1\}}\|=\left(\sum_{i=1}^{2m}\left(G^{\{m,1\}}_{i}\right)^2\right)^{1/2}
\end{align}
Remarks:
\begin{itemize}
\item  The inclusion of the pseudoscalar $G^{\{m,2m\}}$ leads to an additional
  dimension. We have
\begin{align}
&\lambda =\frac{1}{2^m}\left(1\pm \|\tilde{G}\|\right)\\[2ex]
&\text{where}\nonumber\\[2ex]
&\|\tilde{G}\|=\left(\sum_{i=1}^{2m}\left(G^{\{m,1\}}_{i}\right)^2\,+\,\left(G^{\{m,2m\}}\right)^2\right)^{1/2}.
\end{align}
\item For pure states the parameter domains are ,of course,
\begin{align}
&\text{The (2m-1)-sphere $\|G^{\{m,1\}}\|=1$ for vector configurations}\\[1ex]
&\text{and the 2m-sphere $\|\tilde{G}\|=1$ for the pseudoscalar$+$vector}\nonumber\\
&\text{configuration}.
\end{align}
\item Mixed states are represented by the corresponding spheres with radius
  $\|\tilde{G}\|<1$. 
\end{itemize}
\item\underline{2-Tensor configurations}: $k_{tensor}=2$\\[1ex]
Probability spectra turn out to be $2^{m-2}$-fold degenerate: a spectrum is
built up by one quartet repeated $2^{m-2}$ times.We express these four
eigenvalues in terms of $O\left(2m\right)$-invariants. In
the following we shall present explicit calculations for the cases $m=2$ and
$m=3$ and then generalize our findings to the general case.    
\begin{itemize}
\item {\bf $m=2$} : The eigenvalues are
\begin{align}
&\lambda=\frac{1}{4}\left(1\pm\sqrt{r\pm\sqrt{2r^{2}-T_4}}\right)\\[2ex]
&\text{where}\nonumber\\[2ex]
&r=\frac{1}{2}\text{trace}\left(\left(G^{\{m,2\}}\right)^{T}.
  G^{\{m,2\}}\right)\quad\left(\text{Frobenius norm}\right)^2\\[2ex]
&T_4=\text{trace}\left(\left({G^{\{m,2\}}}^{T}.
    G^{\{m,2\}}\right)^2\right).
\end{align}
We see that the eigenvalues depend on only two invariants $r$ and $T_4$
$\footnote{In the definition of the Frobenius norm we include, because of
  (anti-)symmetry, a factor of $1/2$ : $r=\sum_{i<j}^{2m}m_{ij}^2$ where
  ($m_{ij}$) is a $2m\times 2m$ (anti-)symmetric matrix.}$.
\item {\bf $m=3$} :
For $m\geq 3$ new invariants appear (see the discussion at the end of this
section), the characteristic polynomial $P^{8}$ can be shown to factorize into
2 polynomials $P_{4,\pm}$ of degree $4$ which differ by the sign of
$D^{\left(3\right)}$.  
\begin{align}
P_{4,\pm}\left(z\right)=&z^4\;-\;z^3/2\;+\;\left(3-r\right)z^{2}/32\;+\nonumber\\[2ex]
&\left(r-1\pm\frac{64}{3}\,D^{\left(3\right)}\right)z/128\;+\nonumber\\[2ex]
&\left(2-\left(r+1\right)^2+T_4\mp\frac{256}{3}D^{\left(3\right)}\right)/4096\\[2ex]
&\text{where}\nonumber\\[2ex]
&D^{\left(3\right)}=\epsilon_{i_1,i_2,i_3,i_4,i_5,i_6}
\,G^{\{m,2\}}_{i_1,i_2}\,G^{\{m,2\}}_{i_3,i_4}\,G^{\{m,2\}}_{i_5,i_6}
\end{align}
(as usual repeated indices are summed over).\\
The eigenvalues of $P_{4,\pm}$ are even in
$D^{\left(3\right)}$ and depend only on
$\left(D^{\left(3\right)}\right)^2$:\\[1ex]
\underline{the octet of
eigenvalues therefore is degenerate in 2 quartets}.\\[3ex]
Under the assumption that the 2-tensor configuration is such that
$D^{\left(3\right)}$ vanishes we again find the $m=2$ relation
\begin{align*}
&\lambda=\frac{1}{8}\left(1\pm\sqrt{r\pm\sqrt{2r^{2}-T_4}}\right)\\[2ex]
&\text{if }\, D^{\left(3\right)}=0\\[2ex]
&r\text{ is the Frobenius norm and }\\[2ex]
&T_4\text{ is the trace invariant of scale
  dimension 4 defined above.}
\end{align*}
The degeneracy into 2 quartets is explicitly seen in this case.
\item {\bf $m=m_0\geq 4$}: At this stage of affairs the following 'Vermutung'
  is plausible:\\ A $k_{tensor}=k_0\leq m_0$ configuration is $2^{m_0-k_0}$-fold
  degenerate and consists of $2^{m_0-k_0}$ $2^{k_0}$-plets. Algebraic
  solutions of the spectral decomposition can be found for $k_0\leq
  2$ and \underline{all} $m$. A direct though not particularly elegant proof
  of this 'Vermutung' is possible by calculating the characteristic polynomial
  of the corresponding tensor configuration using e.g. the relation
  $\text{Det}A=e^{\text{trace}\left(\log A\right)}$. For example it is easily
  seen that for vector configurations $k_{tensor}=1$ the characteristic
  polynomial $P^{\{m\}}$ factorizes as proposed for all $m$
\[
P^{\{m\}}=\left(\lambda^2-\frac{\lambda}{2^{\left(m-1\right)}}+\frac{\left(1-r\right)}{2^{2m}}\right)^{2^{m-1}}
\]
  For 2-tensor configurations we obtain
\begin{align*}
&\text{Defining}\\[1ex]
&\bar{P}_{\pm}:=z^4-4z^3+2\left(3-r\right)z^2-\left(r-1\pm\frac{64}{3} D^{\left(3\right)}\right)z+\\[1ex]
&\left(2-\left(r+1\right)^{2}\mp\frac{256}{3} D^{\left(3\right)}+T_4\right)\\[2ex]
&\text{we find}\\[1ex]
&P^{\{m\}}\left(\lambda\right)=\left(\bar{P}_{+}\bar{P}_{-}\right)^{2^{m-3}}|_{z=2^{m}\lambda}
\end{align*}
                                                                                                                        
  I shall not spell out the not very inspiring details.\\
  Anticipating the discussion proposed in the next
  paragraph we shall   sketch a proof that the occurrence of a third order
  invariant   $D^{\left(3\right)}$ is possible for $m\geq 4$. For $m=3$ the
  $O\left(2m\right)$ 2-tensor allows for the construction of an
  anti-symmetric 2-tensor of scale dimension 2 given two 2-tensors
\begin{equation}
\tilde{A}_{i_1,i_2}=\epsilon_{i_1 ,i_2 ,i_3 ,i_4 ,i_5 ,i_6}G^{\{3,2\}}_{i_3
  ,i_4}G^{\{3,2\}}_{i_5 ,i_6}
\end{equation}
and therefore of the invariant
\[
D^{\left(3\right)}=\text{trace}\left(\tilde{A}.G^{\{3,2\}}\right);
\]
for $m\geq 4$ higher tensors $G$ are required to be contracted to third order
invariants (i.e. scale dimension=3 (see below)). We see that the 
roots of the 4-th order polynomials - 4-th order because of the degeneracy of
the 2-tensor configurations described above - are functions of the two even
invariants $r$ and $T_4$ defined in (32) and (33) and a third order
invariant. The eigenvalues are expected to be given by (31) for all
$G^{\{m,2\}}$ under the condition $D^{\left(3\right)}=0$ constructed in the way discussed. 
\item The parameter domains for 2-tensor configurations are now determined in
  a straightforward manner. In a ($r-T_4$)-diagram positivity and normalisation
  leads to the inequality
\begin{equation}  
\text{max}\left(\left(r+1\right)^2-2,0\right)\leq T_{4}\leq 2r^2\quad0\leq
r\leq 1
\end{equation}
  for the admissible $r,T_4$ values. Figure 1 shows the corresponding
  diagram. 
\begin{figure}
\centering\epsfig{figure=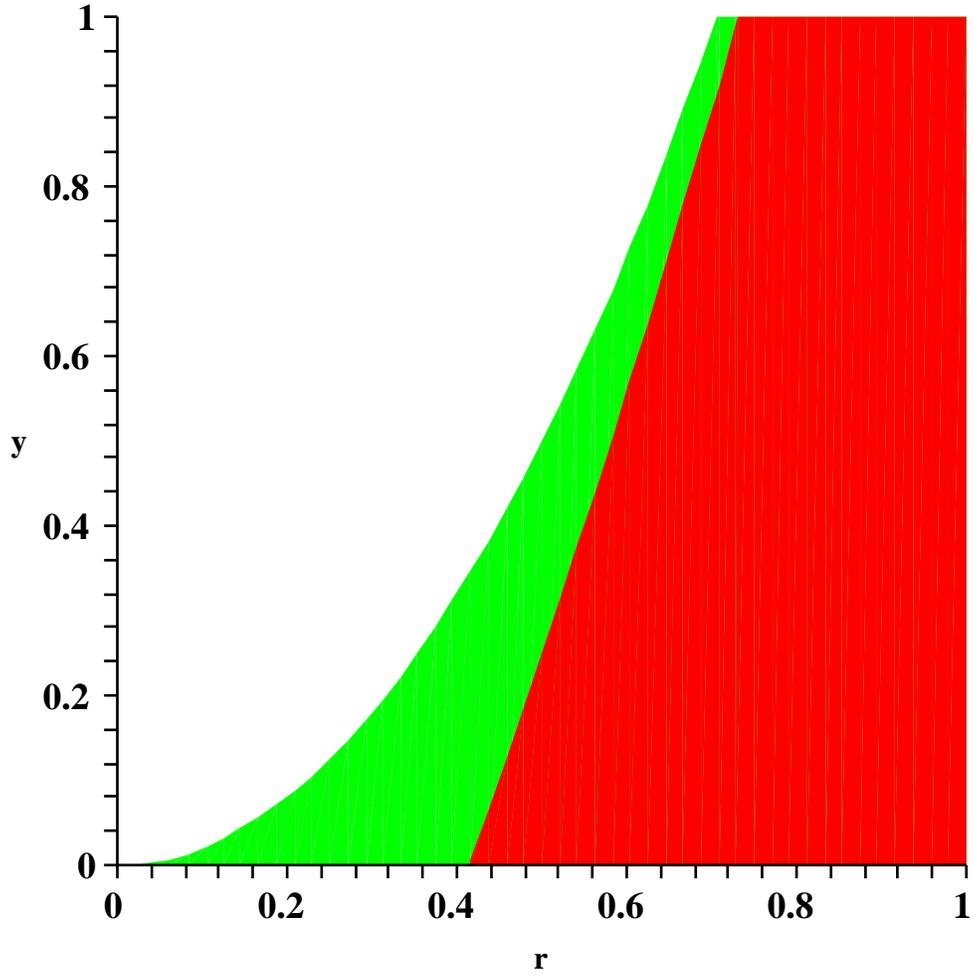,height=12.cm,width=12.cm,bburx=12.cm,bbury=12.cm}
\caption{The admissible 2-tensor domain in a $\left(r,T_4\right)$-plot.}
\end{figure}

It is more convenient to introduce new variables
\[
\left(r,T_4\right)\longrightarrow\left(r,z=\frac{1}{2}-\sqrt{2r^2-T_4}\right),
\]
  the inequalities (37) now read
\begin{align*}
&\frac{1}{2}-z\leq r\leq \frac{1}{2}+z\quad 0\leq z\leq \frac{1}{2}\\[1ex]
&\frac{1}{2}+z\leq r\leq \frac{1}{2}-z\quad -\frac{1}{2}\leq z\leq 0.
\end{align*}
  In terms of the $N_m:=m\left(2m-1\right)$ matrix elements of
  $G^{\{m,2\}}$ the invariants $r,T_4$ or $r,z$ correspond to the following
  geometrical 'balls' 
\begin{align}
&\text{For all $m$ we have}\nonumber\\[1ex]
&r=\sum_{i<j}^{2m}\left(G^{\{m,2\}}_{i,j}\right)^2\nonumber\\[1ex]
&\text{ i.e. the $\left(N_m -1\right)$-sphere of the Frobenius norm}\\[1ex]
&\text{For $m=2$}:\nonumber\\[1ex]
&\text{a straightforward calculation shows}\nonumber\\
&2r^2-T_{4}=\frac{1}{16}\text{trace}\left(\tilde{G}^{\{2,2\}}.G^{\{2,2\}}\right)^2\\[1ex]
&\text{where}\nonumber\\[1ex]
&\tilde{G}^{\{2,2\}}\text{ is the dual tensor }\;\tilde{G}^{\{2,2\}}_{i_1 ,i_2}=\epsilon_{i_1 ,i_2 ,i_3
  ,i_4}G^{\{2,2\}}_{i_3 ,i_4}.\nonumber
\end{align}
($\tilde{A}:=\tilde{G}^{\{3,2\}}\text{ the tensor dual to }G^{\{3,2\}}$)
\begin{align}
&\text{For $m=3$ we find}\nonumber\\[1ex]
&2r^2-T_4=\frac{1}{16}\|\tilde{A}\|^2=\frac{1}{32}\text{trace}\left(\tilde{A}^T\tilde{A}\right)\\[1ex]
&\|\tilde{A}\|\text{ is the Frobenius norm of the 2-tensor (32).}\nonumber
\end{align}  
\end{itemize}
The parameters $a_{i,k}:=G^{\{m,2\}}_{i,k}$ are seen to lie in generalized elliptic
'tunnel' domains (see below). In detail we have
\begin{itemize}
\item $m=2$:\\ Expressing $z$ in terms of $a_{i,k}$
\begin{align}
&z=\frac{1}{2}-2\,|a_{1,2}a_{3,4}+a_{1,3}a_{4,2}+a_{1,4}a_{2,3}|\\[1ex]
&\text{remember}
\quad\lambda=\frac{1}{4}\left(1\pm\sqrt{r\pm\left(\frac{1}{2}-z\right)}\right)\nonumber
\end{align} 
we shall see that the admissible values $-\frac{1}{2}\leq z\leq \frac{1}{2}$ lie in
a generalized elliptic 'tunnel' domain embedded in $\mathbb{R}^6$.
\item $m=3$:\\ The analogeous expression reads in this case
\begin{align}
z=&\frac{1}{2}-2\left(\left(a_{4,5}a_{3,6}-a_{4,6}a_{3,5}+a_{3,4}a_{5,6}\right)^2\right.\\[1ex]
&\left.+\left(a_{2,5}a_{4,6}-a_{4,5}a_{2,6}+a_{2,4}a_{3,6}\right)^2\right.\nonumber\\[1ex]
&\left.+\text{ thirteen similar terms }\right)^{1/2}\nonumber
\end{align}
and $z=z\left(a_{i,k}\right)$ is the 'tunnel' domain embedded
in $\mathbb{R}^{15}$ which we shall illustrate at the end of the section.
\end{itemize} 
\item I should include a short discussion of a qualitative method for the
  construction of $O\left(2m\right)$-invariants by inference. First of all we
  assign the scale dimension $\delta=1$ to the tensor $G^{\{m,k\}}\text{ all
  }m,k$. The eigenvalues then have $\delta=1$, the characteristic polynomial
  $P\left(z\right)=\sum_{i=0}^{d} c_{i}z^i$ of degree $d$ has $\delta=d$, the coefficients
  have $\delta^{d-i}$. Hence $c_{i}$ is composed of invariants of scale
  dimension $\leq d-i$ (counting dimensions such that by putting $G^{\{m,0\}}=1$
  (and thus fulfilling the normalisation condition) we mean that unity carries
  one dimensional unit). In detail we have the following invariants
\begin{itemize}
\item $\delta=d=4$ :
$T_4$ and $r^2$.\\[1ex] We reiterate the identities introduced above.\\ $m=2$:
\begin{align}
2r^2-T_4&=\frac{1}{16}\left(\epsilon_{i_1 ,i_2 ,i_3 ,i_4}G^{\{2,2\}}_{i_1
  ,i_2}G^{\{2,2\}}_{i_3 ,i_4}\right)^2\nonumber\\[1ex]
&=4\,\text{Determinant}\left(G^{\{2,2\}}\right)
\end{align}
m=3: 
\begin{align}
&2r^2-T_4=\nonumber\\[1ex]
&\frac{1}{32}\epsilon_{i_1 ,i_2 ,i_3 ,i_4 ,i_6 ,i_5}G^{\{3,2\}}_{i_1
  ,i_2}G^{\{3,2\}}_{i_3 ,i_4}\epsilon_{i_5, i_6, i_7 ,i_8 ,i_9 ,i_{10}}G^{\{3,2\}}_{i_7
  ,i_8}G^{\{3,2\}}_{i_9 ,i_{10}}
\end{align}
\item $\delta=3$ :
The invariants are the $D^{\left(3\right)}$ discussed above.
\item $\delta=2$ :
The only invariant is $r$ defined above.
\item $\delta=1$ : $G^{\{m,0\}}$\\[1ex]
  Normalisation forces the only invariant, the scalar $G^{\{m,0\}}=1$, to
  be counted with $\delta =1$. The term of order $\lambda^0$, the invariant
  $D^{\left(3\right)}$, should be read as $\left(D^{\left(3\right)}.G^{\{3,0\}}\right)$.
\end{itemize} 
\item Visualisation of 'tunnel' domains:\\[1ex] We now now illustrate the
  domains for the matrix elements $G^{\{m,2\}}_{i,k}$ prescribed by the
  probability interpretation of the eigenvalues (31) (as a reminder, these
  formulae hold exactly for $m=2$ and for $m\geq 3$ when we demand certain scalar
  products of pseudo-tensor(vector) configurations vanish,
  $D^{\left(3\right)}$ for $m=3$). For obvious reasons we restrict the
  configurations to three non-vanishing matrix elements, e.g.
\begin{align*}
&G^{\{m,2\}}_{1,2}:=x\quad G^{\{m,2\}}_{3,4}:=y\quad
 G^{\{m,2\}}_{2,3}:=z\\[1ex]
&G^{\{m,2\}}_{i,k}:=0\quad\text{otherwise}.
\end{align*}
The eigenvalues then are
\begin{align}
&\lambda_{1,2}=\frac{1}{4}\left(1\pm\alpha_{+}\right)\\[1ex]
&\lambda_{3,4}=\frac{1}{4}\left(1\pm\alpha_{-}\right)\\[1ex]
&\text{where}\nonumber\\[1ex]
&\alpha_{\pm}=\sqrt{\left(x\pm y\right)^2+z^2}.
\end{align}  
The domains are determined by the inequalities
\[
0\leq\alpha_{+}\leq 1\;\;\wedge\;\;0\leq\alpha_{-}\leq 1;
\]
the admissible domains have to be subsets of these parameter regions which
graphically represent two 'orthogonal' 'tunnels' with symmetry axes $y=\pm x$
and elliptic cross sections, half-axes $1\text{ and }\sqrt{\frac{1}{2}}$ as depicted in Figure
2.
\begin{figure}
\centering\epsfig{figure=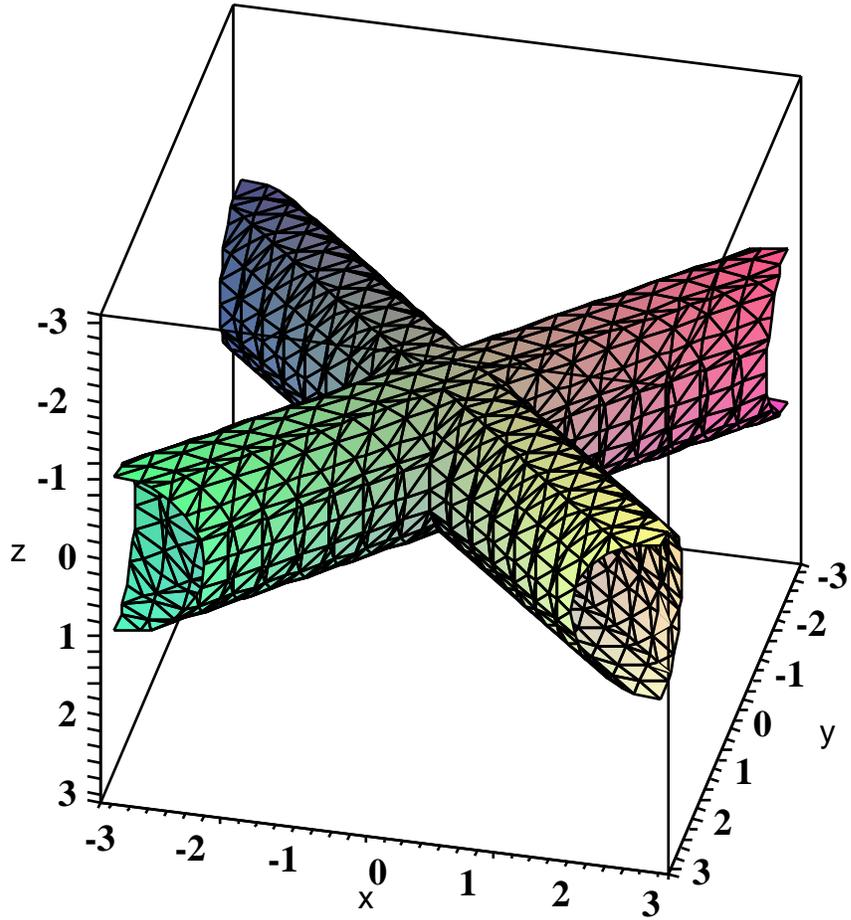,height=12.cm,width=12.cm,bburx=12.cm,bbury=12.cm}
\caption{The domains (47) as a function of unrestricted $\{x,y,z\}$}
\end{figure}
The physical domain is finally constructed as the intersection
$\mathfrak{Int}=\text{tunnel}_{\alpha_{+}}\cap\text{tunnel}_{\alpha_{-}}\cap\{[x,y,z]|0\leq
x,y,z\leq
1\}$ where the last set, the cube with edges
$[x_0,y_0,z_0],\;\;x_0,y_0,z_0=0,1$ represents the positivity condition, the
correct normalisation is guaranteed by (45) and  (46). Figure 3 illustrates
this intersection. 
\begin{figure}
\centering\epsfig{figure=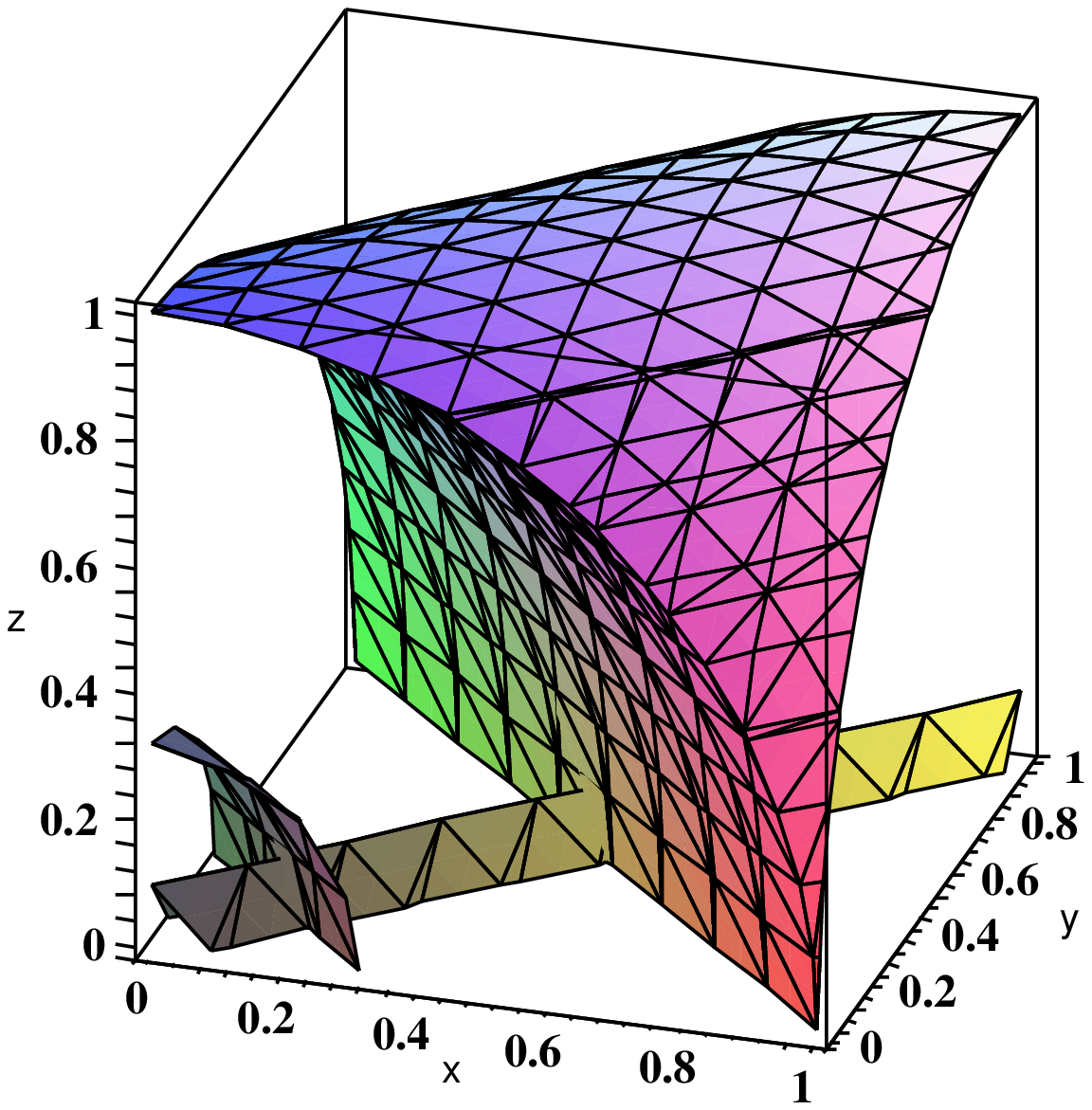,height=12.cm,width=12.cm,bburx=12.cm,bbury=12.cm}
\caption{The intersection $\mathfrak{Int}:$ surfaces $\alpha_{+}=1.\text{
    and }\,0.1$ ;  $\,\alpha_{-}=1.\text{ and }\,0.01$ are depicted.}
\end{figure}     
\end{itemize}

\section{An Alternative Parameter Classification}

In the following I shall describe a classification which has the charm of
accommodating a larger set of parameters into one tensor representation but
is algebraically incomplete. Whether this has disadvantages when it comes to
physical applications can be decided only after a clarification of the r\^ole
of discrete transformations (similar to Parity and Charge conjugation in Dirac
theory). We postpone such questions and proceed as follows.\\ Put
\[
\check{\Gamma}^{\{m,1\}}:=[\,\hat{\Gamma}^{\{m,1\}}_1,\ldots,\hat{\Gamma}^{\{m,1\}}_{2m},\Gamma^{\{m\}}_{2m+1}\,].
\]
Then as already stated the $\check{\Gamma}^{\{m,1\}}_{i}$
fulfil the anti-commutation relations
\begin{align}
&\check{\Gamma}^{\{m,1\}}_{i}.\check{\Gamma}^{\{m,1\}}_{j}+
\check{\Gamma}^{\{m,1\}}_{j}.\check{\Gamma}^{\{m,1\}}_{i}=2\,\delta_{i,j}\mathbb{I}\\[1ex]
&i,j=1\ldots 2m+1.\nonumber 
\end{align}
Following equations (5) and (6) we write
\[
\check{\Gamma}^{\{m,k\}}_{i_1\ldots i_k}=F_{Norm}^{\{m,k\}}\;\left(\varepsilon_{i_1\ldots
  i_{2m}}\;\check{\Gamma}^{\{m\}}_{i_{k+1}}.\ldots\check{\Gamma}^{\{m\}}_{2m}\right).
\]
The $2^m-1$ parameters of an m-Qubit are then accommodated in the expansion
\begin{align*}
&\varrho^{m-Qubit}=\sum_{k=0}^{m}\check{G}^{\{m,k\}}\circ\check{\Gamma}^{\{m,k\}}\\[1ex]
&\text{where it is essential to note}\\[1ex]
&\sum_{k=0}^{m}\binom{2m+1}{k}=2^{2m}\\[1ex]
&\text{the number of matrix elements defining the state $\varrho$.}
\end{align*}
The point to keep in mind is of course that this expansion is
incomplete. However depending on the r\^ole of the already mentioned 'P'-,
'C'-transformations duality relations among the $\{m,k_0\}$ and
$\{m,2m-k_0\}$ tensors will resolve this problem.\\ This scheme takes care of
  a larger number of parameters
\begin{align*}
&{\bf m=2:}\\[1ex]
&\text{Number of parameters }=
\begin{cases}
5\quad\qquad\,\left(4\right)\,\,&\text{ for }k=1\quad\text{ Vector}\\
5+10\quad\left(4+6\right)\,\,&\text{ for }k=2\quad\text{2-Tensor}
\end{cases}\\[2ex]  
&{\bf m=3:}\\[1ex]
&\text{Number of parameters }=
\begin{cases}
7\quad\qquad\,\left(6\right)\,\,&\text{ for }k=1\quad\text{ Vector}\\
7+21\quad\left(6+15\right)\,\,&\text{ for }k=2\quad\text{ 2-Tensor}
\end{cases}
\end{align*}
We now calculate the vector and 2-tensor spectra for this new
representation:\\[2ex]
for k=1 the problem is already solved, see (38) and (39)\\[2ex]
for k=2 we have\\[1ex]
m=2: the formulae (31)
\[
\lambda=\frac{1}{4}\left(1\pm\sqrt{r\pm\sqrt{2r^{2}-T_4}}\right),
\]
as well as (32) and (33) hold with the replacement
$G^{\{m,2\}}\longrightarrow\check{G}^{\{m,2\}}$\\[1ex]
m=3: The same formulae hold if we replace the condition
\begin{equation} 
D^{\left(3\right)}=0
\end{equation}
by 
\begin{align}
&\text{the $\;O\left(2m+1\right)$-invariant }\;\sum_{i=1}^{6}V_i^2=0\\[1ex]
&\text{i.e. }V_i=0\quad i=1\ldots 7\nonumber\\[1ex]
&\text{with}\nonumber\\[1ex]
&V_i=\epsilon_{i,i_1,\ldots,i_6}\check{G}^{\{3,2\}}_{i_1,i_2}
\check{G}^{\{3,2\}}_{i_3,i_4}\check{G}^{\{3,2\}}_{i_5,i_6}.         
\end{align}
Note that $V_7=D^{\left(3\right)}$. For $m\geq 4$ the situation is a bit more
involved. The corresponding maps (sub-'pseudovectors')
$\footnote{$\mathbb{M}\left(\mathbb{R},2m\right)$ is the space of $2m\times 2m$
real matrices}$
\[
\mathbb{M}\left(\mathbb{R},2m\right)_{i}\rightarrow\mathbb{R}\quad i=1\ldots
2m+1
\]
 of scale dimension $m$ constructed analogously to (51):
\[
V_{i_1}=\epsilon_{i_1,\ldots ,i_{2m+1}}\;\,\check{G}^{\{m,2\}}_{i_2,i_3}\ldots
\check{G}^{\{m,2\}}_{i_{2m},i_{2m+1}}
\]
carry too high dimension and play no r\^ole on the 2-tensor level. Suitable
'pseudo-tensors' have to be constructed and contracted to (pseudo-)scalars of
the required dimension $6$. The normalisation of states is fixed once and for ever
by normalising the scalar term, positivity is guaranteed by the same
inequalities among now $O\left(2m+1\right)$-invariants obtained above.

\section{Summary}

Given a hermitian matrix with unit trace the decision whether or not it is a
state is not at all trivial. More precisely speaking the a priori construction
of a matrix representation of a state, a density matrix, is non-trivial. The
classic way to determine the eigenvalues of this matrix as a function of its
matrix elements, to solve the characteristic equation, is in general feasible
(by radicals and algebraic operations) only for dimension$\leq 4$. For higher
dimensions e.g. the Descartes' rule can be applied to derive admissible
parameter domains. Doublet ($m=1$), quartet ($m=2$), and eventually octet
($m=3$) structures in m-Qubit spectra can be handled in this way with tolerable effort.
Therefore a systematic study of generacy structures in m-Qubit spectra seems essential.\\ 
The key of the approach we followed is to embed m-Qubit states in Clifford
algebras $\mathfrak{Cl}_{2m}$. The construction of a
basis of this algebra from Clifford numbers obeying the anti-commutation rules
(1) and (48) leads, considering the dual representation (9) of the algebra, to
a classification of states as $O\left(2m\right)$- or
$O\left(2m+1\right)$-tensors; the eigenvalues of these states are functions of
$O\left(2m\right)$- and $O\left(2m+1\right)$-invariants. The number of
parameters controlling positivity is thus considerably reduced. For m-Qubits
the case of degeneracy into doublets leads to a vector classification:
state-parameters lie on $\left(2m-1\right)$- or $2m$-Bloch spheres. The
degeneracy into quartets leads to more involved structures: relations among
invariants and their embedding in parameter spaces are dicussed in some
detail. $m\left(2m\mp 1\right)$-dimensional intersections of 
 'tunnel'-like objects with elliptic cross-sections appear as generalisations of Bloch
  spheres. Progressing to tensors with $k_0\geq 3$ one immediately encounters the obstacle of
  not explicitly knowing the eigenvalues as functions of
  $O\left(2m\left(+1\right)\right)$-invariants, the already mentioned
  Descartes rule then comes into play.\\ The use of direct products of Qubit
  states as basis in m-Qubit state spaces has been proven useful for the
  development of criteria of   separability, see e.g. \cite{asch},
  \cite{jae}, \cite{li}, \cite{ma}, \cite{hi}. Of particular interest in the
  present context are \cite{by}, \cite{ki} and references cited in these
  papers. There the basis $\{B_i\}$ is chosen as the generators of
  $SU\left(2^m\right)$, domains of admissible 'coherence vectors' guaranteeing
  positivity of the density matrix of an m-Qubit are given in terms of Casimir
  invariants. In particular cases degeneracies were detected and the
  corresponding local unitaries described. Deriving explicit domains within
  this formalism soon encounters   considerable problems (notwithstanding the
  essential structural   clarifications gained): to determine the admissible
  domain for a m-Qubit one   obtains $2^{m}-1$ polynomial inequalities with
  maximal degree $2^m$: writing the characteristic polynomial as
  $P\left(\lambda\right)=\sum_{i=1}^{2^m}\left(-1\right)^i a_i \lambda^i\quad$ $a_{2^m}$ is of scale
  dimension $2^m$ in the 'coherence vector' $\vec{n}$, (scale dimemsion($a_i$)$=i$); the
  necessary and sufficient condition for positivity of the density matrix
  $a_i>0\text{ for all i },\;\;a_i=a_i\left(\vec{n}\right)$.\\ The approach we
  follow leads to a $O\left(2m\left(+1\right)\right)$-tensor classification
  and the corresponding degeneracy patterns. Domains of admissible parameters
  can thus be derived for \underline{all} $m$ with increasing complexity for
  increasing order of the $O\left(2m\left(+1\right)\right)$-tensors (the
$k_{tensor}=1,2,3$ cases can be comfortably handled on a standard laptop).

\end{document}